# Microscopic Origin of Interfacial Dzyaloshinskii-Moriya Interaction

Sanghoon Kim[1†★], Kohei Ueda[1,2†], Gyungchoon Go[3], Peong-Hwa Jang[3], Kyung-Jin Lee[3,4], Abderrezak Belabbes[5], Aurelien Manchon[5], Motohiro Suzuki[6], Yoshinori Kotani[6], Tetsuya Nakamura[6], Kohji Nakamura[7], Tomohiro Koyama[8], Daichi Chiba[8], Kihiro Yamada[1], Duck-Ho Kim[1], Takahiro Moriyama[1], Kab-Jin Kim[1,9] & Teruo Ono[1,10★]

[1]*Institute for Chemical Research, Kyoto University, Gokasho, Uji, Kyoto 611-0011, Japan*
[2]*Department of Materials Science and Engineering, Massachusetts Institute of Technology, Cambridge, Massachusetts 02139, USA*
[3]*Department of Materials Science & Engineering, Korea University, Seoul 02841, Korea*
[4]*KU-KIST Graduate School of Converging Science and Technology, Korea University, Seoul 02841, Korea*
[5]*Physical Science and Engineering Division, King Abdullah University of Science and Technology (KAUST), Thuwal 23955-6900, Saudi Arabia*
[6]*Japan Synchrotron Radiation Research Institute (JASRI), Sayo, Hyogo 679-5198, Japan*
[7]*Department of Physics Engineering, Mie University, Tsu, Mie 514-8507, Japan*
[8]*Department of Applied Physics, Faculty of Engineering, The University of Tokyo, Bunkyo, Tokyo 113-8656, Japan*
[9]*Department of Physics, Korea Advanced Institute of Science and Technology, Daejeon 34141, Korea*
[10]*Center for Spintronics Research Network (CSRN), Graduate School of Engineering Science, Osaka University, Osaka 560-8531, Japan*

[†] These authors contributed equally to this work.

★ Correspondence to: makuny80@gmail.com, ono@scl.kyoto-u.ac.jp

**Chiral spin textures at the interface between ferromagnetic and heavy nonmagnetic metals, such as Néel-type domain walls and skyrmions, have been studied intensively because of their great potential for future nanomagnetic devices. The Dyzaloshinskii-Moriya interaction (DMI) is an essential phenomenon for the formation of such chiral spin textures. In spite of recent theoretical progress aiming at understanding the microscopic origin of the DMI, an experimental investigation unravelling the physics at stake is still required. Here, we experimentally demonstrate the close correlation of the DMI with the anisotropy of the orbital magnetic moment and with the magnetic dipole moment of the ferromagnetic metal. The density functional theory and the tight-binding model calculations reveal that asymmetric electron occupation in orbitals gives rise to this correlation.**

Chiral interaction between two atomic spins owing to a strong spin-orbit coupling, which is known as the Dzyaloshinskii-Moriya interaction (DMI), has attracted intense interest [1,2]. In particular, it has been demonstrated that the DMI at the interface between ferromagnetic (FM) and nonmagnetic heavy metals (HM) plays a major role for the formation of chiral spin textures, such as skyrmions [3, 4] and homochiral Néel-type domain walls (DW) [5-7], which are attractive for the development of future information storage technology [8]. Understanding the microscopic origin of the DMI is indispensable for the realization of such chiral spin textures [9,10]. It has been reported that the proximity-induced magnetic moment in HM layers has a critical role for emerging the DMI [11]. However, this proximity effect is still controversial because it has been also reported that the induced magnetic moment has no direct correlation with the DMI in the case of the Co/Pt system [12,13]. On the other hands, theories have predicted that spin-orbit coupling combined with inversion symmetry breaking (ISB) naturally introduces a chirality to conduction electron spins in equilibrium and the interfacial DMI at an FM/HM interface is related to this spin chirality [14-18]. It has also been reported that the spin chirality is a manifestation of the chirality of the orbital magnetism in strongly spin-orbit coupled systems with ISB [19,20]. These previous studies suggest a possible microscopic origin of the interfacial DMI, which has remained experimentally-unaddressed so far.

Here we discuss the microscopic origin of the interfacial DMI with experimental and theoretical studies as follows: First, we show the temperature dependence of the DMI for a Pt/Co/MgO trilayer, which is one of the standard structures used for the studies of the DMI [7,12,21], using the extended droplet model [22]. We find that the DMI increases with decreasing temperature in a range from 300 to 100 K. In general, the electron-phonon interaction promotes thermally-induced hopping between nearest neighbours when increasing the temperature [23]. As a result, it is expected that the difference between in-plane and out-of-plane hopping energies is reduced upon temperature increase. Therefore, changing the temperature of the system allows for charge redistribution between in-plane and out-of-plane orbitals, while preserving the major electronic state of the trilayer unlike other interface control methods such as an ion irradiation [24] or a thermal annealing technique [25], which may cause a permanent atomic rearrangement and thus induce undesired extrinsic effects. To discuss this temperature dependence of the DMI, that of the spin ($m_s$) and orbital ($m_o$) magnetic moments of Co and Pt is studied by X-ray magnetic circular dichroism (XMCD) spectroscopy. We find that $m_s$ values of Co and Pt show weak temperature dependences. On the other hand, the intra-atomic magnetic dipole moment ($m_D$), which is due to the asymmetric spin density distribution [26,27], shows a strong temperature dependence, suggesting a sizable modification of the charge distribution between the in-plane and the out-of-plane *d*-orbitals under temperature variation. We also find that the out-of-plane orbital moment ($m_o^\perp$) shows large temperature dependence while in-plane orbital moment ($m_o^\parallel$) does not, revealing a close connection between the anisotropy of $m_o$ (orbital anisotropy) and the DMI. The *ab-initio* and the tight-binding model calculations suggest that the ISB-dependent electron hopping, which gives rise to the asymmetric charge distribution at the interface of the FM/HM, is a possible microscopic origin of the correlation between the orbital anisotropy and the DMI.

**Results**

**Temperature dependence of the DMI in the Pt/Co/MgO trilayer**

The temperature dependence of the DMI-induced effective field ($H_{DMI}$) of the Pt (2 nm)/ Co (0.5 nm)/MgO (2 nm) trilayer is determined by measuring the nucleation field ($H_n$) applied along the out-of-plane direction and analysing $H_n$ with the extended droplet model as introduced in Ref. 21. The measured $H_n$ as a function of the in-plane field ($H_x$) shows a threshold arising from the DMI as predicted by the extended droplet model (see also Supplementary Information and Method section for details about the analysis and the measurement). Figures 1a-1d show $H_n/H_{SW}(H_x=0)$ as a function of $H_x/H_K$ at various temperatures (a, $T$ = 300 K, b, 200 K, c, 150 K, and d, 100 K). Here, $H_n$ and $H_x$ are normalized by the switching field at $H_x=0$ [$H_{SW}(H_x=0)$] and by the anisotropy field ($H_K$), respectively. Those normalized values allow us to clearly confirm the DMI-dependent threshold with ruling out the temperature dependent characteristics of $H_K$ and the $H_{SW}(H_x=0)$. The temperature-dependent $H_{DMI}$ can be determined from the best fitting using the extended droplet model; $H_{DMI}$ at 300, 200, 150 and 100 K are 166±50, 245±45, 324±15 and 372±30 mT, respectively (Fig. 1e). The temperature-dependent DMI energy density ($D$) is readily calculated from $H_{DMI}$ and $\Delta$ via $H_{DMI} = D/\mu_0 M_S \Delta$, where $\mu_0$ is the Bohr magneton, $M_S$ is the saturation magnetization, and $\Delta$ is the domain wall width [6]. $\Delta$ and $M_S$ values in terms of the temperature are listed in Table I. We find that $D$ has a strong temperature dependence as shown in Fig. 1e; $D$ increases by a factor of 2.2 as the temperature decreases from 300 to 100 K.

**Temperature dependence of the proximity-induced magnetic moment of Pt**

To understand the strong temperature dependence of the DMI, we firstly investigate the role of the induced magnetic moment in the Pt layer because it has been suggested as the microscopic origin of the DMI [11]. The temperature dependence of the Pt induced magnetic moment was measured using the X-ray magnetic circular dichroism (XMCD) method, which enables element-specific analyses of spin and orbital magnetism [29,30]. Figure 2a presents the XMCD, and integration of XMCD spectra measured at the Pt $L_{2,3}$ edge. The intensities of XMCD are ~3% of the XAS edge heights. At both the $L_3$ and $L_2$ edges (around 11.57 keV and 13.28 keV, respectively), the integrated XMCD spectra show temperature dependences. In contrast, there is no temperature dependence in XAS spectra (see inset of Fig. 2a). The total magnetic moment ($m_{total}$), which is the sum of an effective spin magnetic moment ($m_s^{eff} = m_s + m_D$) and $m_o$, was estimated by a sum rule calculation [31] (see the details about the sum rule calculation in Supplementary Information). In this study, we use moment values per single hole rather than those per atom because it is difficult to precisely determine the hole number ($n_h$) of Pt in the Pt/Co/MgO. The moment values normalized by $n_h$ are directly obtained from the sum rule formula without considering $n_h$ [31]. The changes in the induced magnetic moments with temperature are small (~15%) and comparable to the error range of the analysis as shown in Fig. 2b. This suggests that there is no clear relevance between the temperature dependences of the proximity-induced magnetic moments of the Pt layer and the DMI in the Pt/Co/MgO system, which is consistent with the first-principles calculations [12,13].

**Correlation between asymmetric orbital structure of the FM layer and the DMI**

In this section, we study various physical quantities associated with the magnetism of Co such as $m_s$, $m_o^{\parallel}$, $m_o^{\perp}$, and $m_D$ to identify the microscopic origin of the temperature-dependent DMI. Details about the sum rule calculation are explained in the Supplementary Information. The XAS and XMCD of the film are measured at 100, 200, and 300 K. Two incident angles ($\theta$=0° and 70° with respect to the film normal) were used to separately estimate $m_o^{\parallel}$ and $m_o^{\perp}$ values using the relation $m_o(\theta) = m_o^{\perp}\cos^2\theta + m_o^{\parallel}\sin^2\theta$ [32]. Figures 3a and 3b are typical XAS and XMCD spectra at the Co $L_{2,3}$ edges obtained at 0° and 300 K. The XMCD spectra show a clear temperature dependence of their intensity at both 0° and 70° as shown in Figs. 3c-3f. In case of the $m_s$, there is a small increase of 20% (from 0.87 $\mu_B/n_h$ to 1.11 $\mu_B/n_h$) as temperature decreases from 300 to 100K (see Fig. 3g), which is much smaller than the strong temperature dependence of the DMI.

In contrast to $m_s$, the magnetic dipole moment $m_D$, which reflects the asymmetric (aspherical) charge distribution between in-plane and out-of-plane *d*-orbitals [26,27], shows a strong temperature dependence. The value of $m_D$ determined from the sum rule analysis increases from 0.014 $\mu_B/n_h$ to 0.094 $\mu_B/n_h$ (Fig. 3g). In addition, the orbital anisotropy also increases as temperature decreases; $m_o^\perp$ increases from 0.058 $\mu_B/n_h$ to 0.080 $\mu_B/n_h$ whereas $m_o^\parallel$ slightly decreases from 0.044 $\mu_B/n_h$ to 0.039 $\mu_B/n_h$ (Fig. 3h). These results imply a correlation of DMI with $m_D$ and orbital anisotropy. Figure 4a shows $m_o^\perp/m_o^\parallel (=[m_o^\perp(T)/m_o^\parallel(T)]/[m_o^\perp(300K)/m_o^\parallel(300K)])$, and $m_D(T)/m_D(300K)$ plots as a function of the normalized $D=D(T)/D(300K)$ where $T$ inside of parentheses is the measurement temperature. The ratio $m_o^\perp/m_o^\parallel$ increases by 53% as $D$ increases, and $m_D$ shows a much stronger correlation with the DMI; $m_D(100K)/m_D(300K) \approx 6.4$. Because both orbital anisotropy and $m_D$ are closely related to the orbital occupation with ISB [26,33,34], these results suggest that the temperature dependence of DMI is governed by the change in asymmetric electron occupation in orbitals as we discuss with the following theoretical studies.

**Theoretical consideration about the microscopic origin of the DMI**

In order to support the insights obtained from our experimental study, we carry out two types of theoretical calculations: a tight-binding model calculation and *ab-initio* calculation based on density function theory (DFT), and the main result is shown in Figs. 4b and 4c. The details of these two complementary theoretical studies can be found in the Supplementary Information.

For the tight-binding model, we extend the trimer model suggested by Kashid *et al*. [10], which contains two magnetic atoms coupled to a spin-orbit coupled non-magnetic ion. Compared to Kashid's trimer model, we add one more orbital ($d_{xz}$) on the nonmagnetic site, enabling the computation of the orbital anisotropy (see details in Supplementary Information). In order to describe the temperature dependence of parameters in our trimer model, we assume the level broadening increases with the temperature range. This broadening can be phenomenologically explained by magnetization fluctuations and the electron-phonon interaction given by atomic vibrations. As this model calculation is too simple, we do not aim to give a quantitative explanation of the experimental data but provide a qualitative understanding of the experimental result. Figure 4b shows that the tight-binding model calculation reproduces qualitatively our experimental observations; both the DMI and $m_o^\perp$ decrease with temperature, while the $m_o^\parallel$ is almost temperature-independent. Note that the $m_o^\perp$ and $m_o^\parallel$ cases are related to electron hopping with and without ISB, respectively (Supplementary Information). In addition, the difference in the spin density distribution ($\Delta$SD) between the in-plane ($xy$) and out-of-plane ($yz$, $xz$) orbital states decreases as temperature increases as shown in the inset in Fig. 4b [35]. This result implies that the spin distribution variation of the orbital states with ISB is key to change of the orbital moment and the DMI, which is in line with the following DFT result.

To examine the correlation between the DMI and the orbital anisotropy in realistic structures, we also perform *ab initio* calculations for Pt(111)/X ultra-thin films, where X is a 3$d$ transition metal (X=V, Cr, Mn, Fe, Co, Ni). Based on the experimental observation, the microscopic origin of the correlation between the DMI and the orbital anisotropy involves the impact of the temperature on 3$d$-orbital magnetization and their electron filling. In this respect, it is instructive to vary the 3$d$ transition metals on the Pt substrate and examine the general chemical trend. Details about this study are also discussed in Supplementary Information. Figure 4c shows the summary of the DFT calculation. Here, changing the overlayer results in modification of the relative alignment between $3d$ and $5d$ orbitals and thereby in a modification of the charge distribution: more hybridization results in less asphericity of the charge distribution between the in-plane and the out-of-plane, as reflected by the change in $m_D$. Our first principles calculations confirm that $m_D$ is governed by the strength of the interfacial hybridization and the filling of the $3d$ orbitals, which is consistent with the phenomenological explanation deduced from the experimental observations. Furthermore, when increasing the interfacial hybridization (e.g. by moving from Mn to Ni overlayer), both DMI and the orbital anisotropy decrease, reproducing the experimental behaviour obtained upon increasing the temperature. Furthermore, both the DMI and the orbital anisotropy follow the same trend in their signs; $D^{\text{tot}}$ and $m_o^\perp/m_o^\parallel$ at the Pt/V interface have negative sign, while those of other interfaces have positive sign. We do not find any correlation between the DMI and the induced magnetic moment of Pt, which is also consistent with the experimental results.

## DISCUSSION

Our experimental study on the temperature dependence of the DMI suggests that the interfacial DMI in FM/HM bilayers originates from the asymmetric charge distribution caused by the ISB, as evidenced by the orbital anisotropy and magnetic dipole moment ( $m_o^\perp/m_o^\parallel$, $m_D$). Our DFT simulation and tight-binding calculation provide a clear evidence of the close link between the DMI and orbital physics. Based on the theoretical discussions, the temperature-dependent $m_D$ and $m_o$ indicate that the increasing in the temperature promotes the phonon-induced electron hopping between the out-of-plane and in-plane orbitals, thereby resulting in a reduced asphericity (or reduced asymmetry) of spin density distribution over the $d$-orbitals and thus quenching $m_D$.

In our experiment, however, the correlation between $m_o^\perp/m_o^\parallel$ and the DMI is only semi-quantitative; i.e., the temperature-dependent change in the DMI does not perfectly scale with the temperature-dependent change in $m_o^\perp/m_o^\parallel$ (see Fig. 4). This semi-quantitative correlation between the DMI and $m_o$ demands a more detailed discussion. In systems with ISB, $m_o^\perp$ can be decomposed into ISB-independent part and ISB-dependent part. Given that ISB is an essential ingredient for the DMI, there should be a direct correlation between the DMI and ISB-dependent $m_o$, as evidenced by our tight-binding model calculation. However, $m_o^\perp$ also has an ISB-independent part, which precludes a direct and quantitative correlation between the DMI and $m_o^\perp$. This statement can be rephrased technically as the DMI involves only off-diagonal elements of the spin-orbit coupling operator [10], while $m_o$ involves all of them. We note, however, that even with this uncertainty, the experimentally observed correlation between the DMI and $m_o$ anisotropy for the Pt/Co/MgO structure is rather clear, implying that the ISB-dependent $m_o$ would dominate over the ISB-independent one in this structure. Therefore, our findings with both experimental and theoretical studies provide a link between orbital physics and spin-orbit related phenomena such as the DMI, which are essential for spin-orbitronic devices.

**Method**

  **Film preparation and device fabrication.** Si/Ta (1.5)/Pt (2)/ Co (0.5)/MgO (2)/HfO (5) (in nm) film with perpendicular magnetic anisotropy was deposited on an undoped Si substrate by DC magnetron sputtering and the atomic layer deposition technique. A 3-μm-wide Hall cross structure were fabricated using the photo lithography and the Ar ion milling. For the XMCD measurement, the same stack film was prepared.

  **Nucleation field measurement.** Angular dependent coercivity of the Co/Pt Hall device was measured to estimate the $H_n$ of the magnetic droplet at 300, 200, 150, and 100K. The angle between magnetic field and the sample normal was varied from 0º to 89º rotating the electromagnet. At each angle, magnetic field was swept with in ±0.5 T to observe the coercivity. Details about this measurement is also discussed in Ref. 18.

  **XMCD measurement; 1) Soft X-ray:** Soft X-ray absorption spectra were measured using the total electron yield method with 96% circularly polarized incident X-rays at the BL25SU at SPring-8. XMCDs at the Co $L_3$ and $L_2$ edges (in a range between 770~840 keV) were recorded in the helicity-switching mode with an applied magnetic field of 1.9 T. Homogeneity of the magnetic field was better than 99% for $\phi$10 mm at the sample position. The incident light direction was inclined by 10° with respect to the magnetic field direction. Temperature was varied from 300K to 100K using a continuous liquid He flow type cryostat. **2) Hard X-ray:** XMCD experiments using hard X-rays were carried out at BL39XU of SPring-8. A circularly polarized X-ray beam with a high degree of circular polarization (> 95%) was produced with a transmission-type diamond X-ray phase retarder of 1.4-mm-thickness. X-ray absorption spectra (XAS) of the film were observed at a 0.6 T magnetic field applied parallel to the X-ray propagation direction of which an incident angle was 0° with respect to the surface normal. The X-ray fluorescence yield mode was used to record the spectra. The X-ray energy was scanned around the Pt $L_3$ and $L_2$ edges in a range between 11.5~13.5 keV, reversing the X-ray photon helicity at 0.5 Hz. In this manner, two helicity dependent spectra $I^+$ and $I^-$ were recorded simultaneously. Here, $I^+$ and $I^-$ denote the intensities when the incident photon momentum and the magnetization vectors are parallel and antiparallel, respectively. The XMCD spectrum, $\Delta I = I^+ - I^-$, is given by the difference of the two spectra. Detailed experimental setups for the soft and hard X-ray MCD measurements are described elsewhere [36, 37].

**Acknowledgements**

We also thank H.-W. Lee for fruitful discussion about the relation between the orbital magnetism and the DMI. This work was partly supported by JSPS KAKENHI Grant Numbers 15H05702, 26870300, 26870304, 26103002, 25220604, Collaborative Research Program of the Institute for Chemical Research, Kyoto University, R & D project for ICT Key Technology of MEXT from the Japan Society for the Promotion of Science (JSPS), and the Cooperative Research Project Program of the Research Institute of Electrical Communication, Tohoku University. This work has also been performed with the approval of the SPring-8 Program Advisory Committee (Proposal Nos. 2015A0117, 2015A0125). S. K. and D.-H.K. was supported from overseas researcher under Postdoctoral Fellowship of Japan Society for the Promotion of Science (Grant Number 2604316, P16314). A.M. and A.B. acknowledge support from King Abdullah University of Science and Technology (KAUST). G.G., P.-H.J. and K.-J. L. also acknowledge support from National Research Foundation of Korea (NRF-2015M3D1A1070465, 2017R1A2B2006119). KJK acknowledges support from the KAIST start-up funding.

**Author contributions**

S. K., K. U., K.-J. K., T.M. and T.O. conceived and designed the study. S.K., K.U., T.K., and D.C. fabricated the device and performed the nucleation field measurement. S.K., D.–H. Kim, P.-H.J., T.M., K.-J.K., and T.O. contributed to determine the DMI-induced effective field using the droplet model. S.K., M.S., Y.K., T.N. and K.–J.K. designed and performed XMCD experiment, and analysed the data. K.U. and K.Y characterizes the magnetic properties of films. A.B., A.M. and K. N. supported theoretical analyses with *ab initio* calculation. G.G. and K.–J.L supported the tight-binding model for the temperature dependence of the DMI and orbital magnetism. All authors discussed the results, and wrote the manuscript.

**Additional Information**

Supplementary Information is available in the online version of the paper. Reprints and permissions information is available at www.nature.com/reprints. Correspondence and requests for materials should be addressed to S. K. and T. O.

**Competing financial interests**

The authors declare no competing financial interests.

**Table I**

**Parameters to estimate the DW energy.** $\Delta$, the domain wall anisotropy ($K_D$), and the Bloch-type DW energy ($\sigma_0$) values obtained from the best fitting of $H_n(H_x)/H_{SW}(H_x=0)$ vs $H_x/H_K$ plots using the extended droplet model. The values are comparable with previous reported values in Ref. 28. $M_s$ values were measured using a superconducting quantum interference device magnetometer.

| Temperature | $M_s$ (T) | $\Delta$ (nm) | $K_D$ ($10^3$ kJ/m$^3$) | $\sigma_0$ (mJ/m$^2$) |
|---|---|---|---|---|
| 300 | 1.06 | 1.95±0.3 | 4.62±0.4 | 8.42±1.0 |
| 200 | 1.22 | 1.74±0.3 | 6.85±2.0 | 11.52±1.5 |
| 150 | 1.28 | 1.55±0.25 | 8.47±2.0 | 13.33±1.5 |
| 100 | 1.33 | 1.55±0.4 | 9.15±2.0 | 12.00±2.0 |

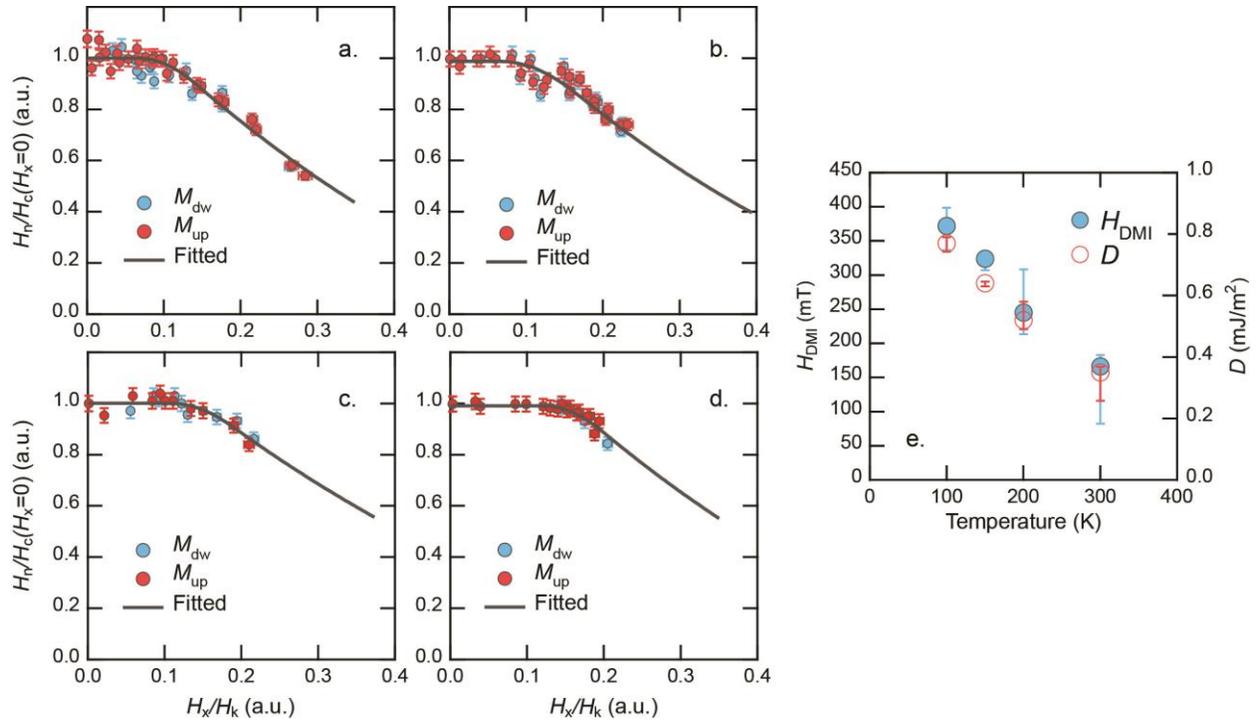

**Figure 1| $H_{DMI}$ measurement from the $H_n$ of the droplet. a., b., c.,** and **d.** The $H_n/H_{SW}(H_x=0)$ vs $H_x/H_K$ plots measured at 300, 200, 150, and 100K, respectively. The grey solid lines are the best fitting results using the droplet model. Here, the vertical axis is normalized by the nucleation field $H_{SW}(H_x=0)$ at $H_x=0$ and the horizontal axis is normalized by the effective perpendicular anisotropy field $H_K$. **e.** Plots of $H_{DMI}$ and $D$ in terms of $T$.

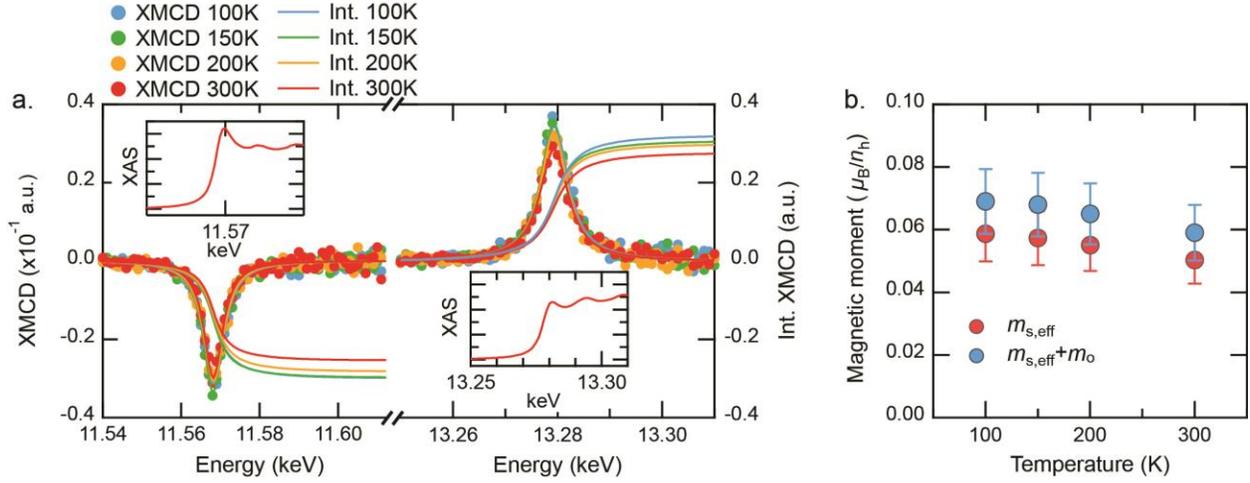

**Figure 2. Temperature dependence of the proximity induced moment in the Pt layer.**
**a.** The XMCD, and integrated XMCD spectra at the Pt $L_3$ and $L_2$ edges in terms of temperature. The insets are the XAS spectra at the Pt $L_3$ and $L_2$ edges. **b.** Temperature dependence of the $m_s^{eff}$ and $m_{total} = m_s^{eff} + m_o$.

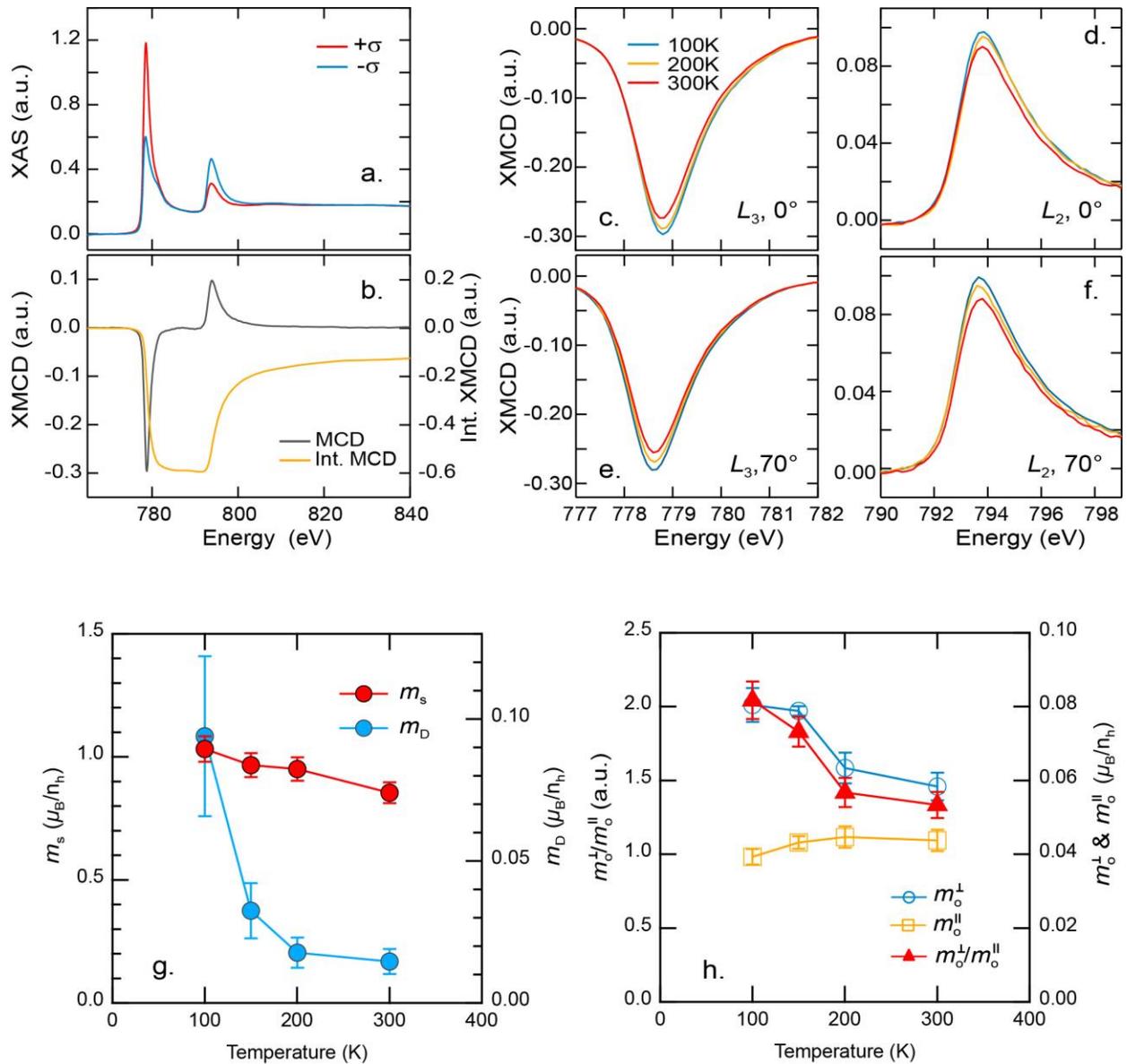

**Figure 3| Temperature dependence of the Co magnetic moments. a.** XAS spectra for positive (σ⁺) and negative (σ⁻) X-ray helicities, **b.** XMCD and integrated XMCD spectra at 0° with 300 K. The temperature dependence of XMCD spectra at the Co $L_3$ and $L_2$ edges measured at (**c.** and **d.**) 0° and (**e.** and **f.**) 70°. **g.** Plots of $m_S$ and $m_D$ versus temperature as a function of the X-ray incident angle. **h.** $m_o^\perp, m_o^\parallel$, and $m_o^\perp/m_o^\parallel$ as functions of temperature.

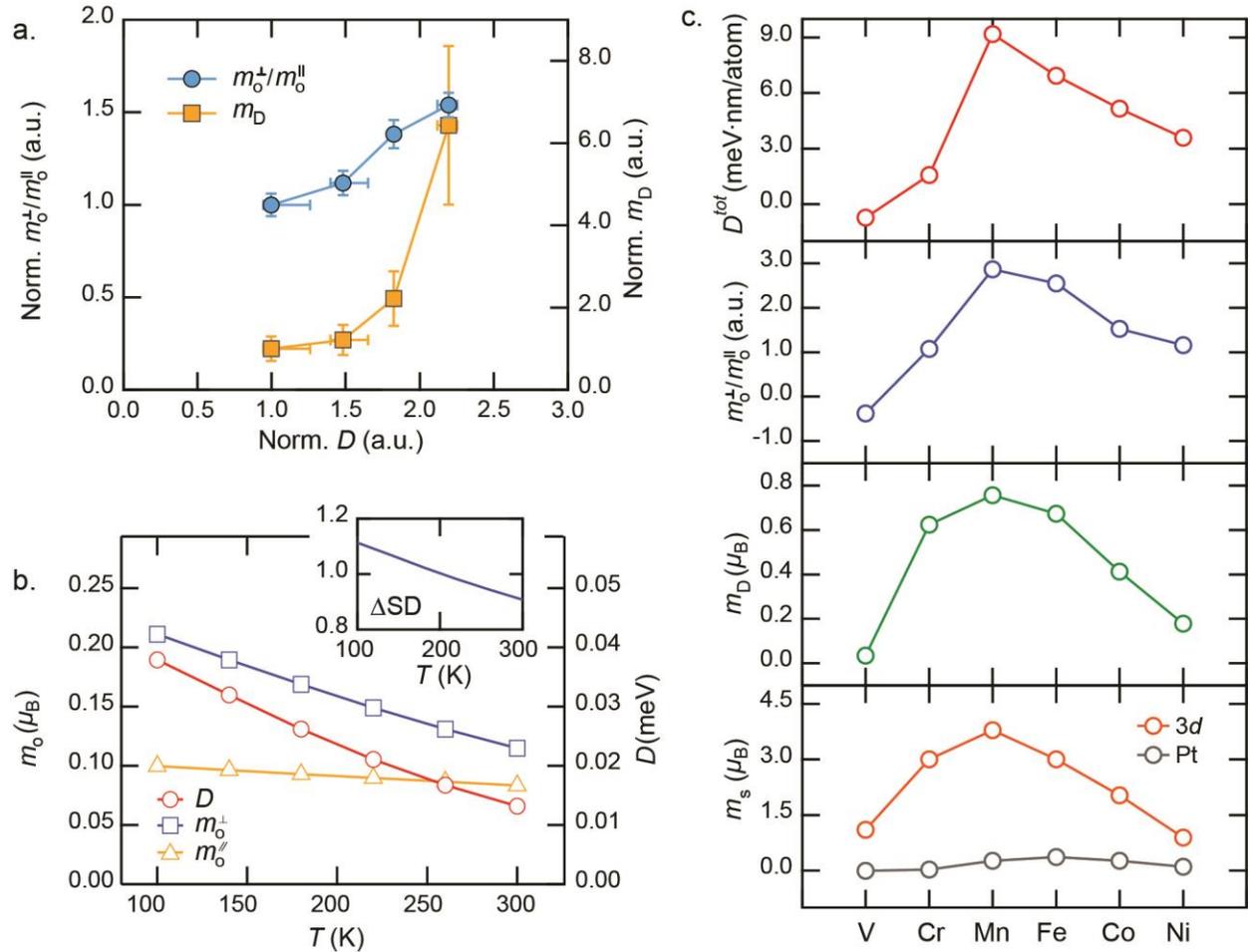

**Figure 4| Correlation of DMI with $m_o^\perp/m_o^\parallel$ and $m_D$, and theoretical calculations based on the tight binding model and DFT. a.** Normalized $m_o^\perp/m_o^\parallel$ and $m_D$ versus normalized $D$. Parameters for all temperatures are normalized by the values measured at 300K. **b.** Calculated $m_o^\perp$, $m_o^\parallel$, and $D$ values based on the tight-binding model as a function of temperature. Inset shows $\Delta$SD between out-plane and in-plane components as a function of temperature. **c.** Physical parameters such as total $D$ ($D^{tot}$), $m_o^\perp/m_o^\parallel$, $m_D$ and $m_s$ obtained by DFT calculation. Strength and sign of $D^{tot}$ are calculated around their magnetic ground state using the combination of the relativistic effect spin-orbit coupling with the spin spirals. A positive sign of $D^{tot}$ indicates a left-rotational sense or "left chirality".